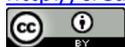

# A Comparison of Classifiers in Performing Speaker Accent Recognition Using MFCCs

## Zichen Ma, Ernest Fokoué

Center for Quality and Applied Statistics, Rochester Institute of Technology, Rochester, USA
Email: zxm7743@rit.edu, epfeqa@rit.edu

## Abstract

**An algorithm involving Mel-Frequency Cepstral Coefficients (MFCCs) is provided to perform signal feature extraction for the task of speaker accent recognition. Then different classifiers are compared based on the MFCC feature. For each signal, the mean vector of MFCC matrix is used as an input vector for pattern recognition. A sample of 330 signals, containing 165 US voice and 165 non-US voice, is analyzed. By comparison, $k$-nearest neighbors yield the highest average test accuracy, after using a cross-validation of size 500, and least time being used in the computation.**

## 1. Introduction

A popular task in signal processing is the classification of different people by their accents. That is, given an input signal, the task is to classify the accent of the speaker [1]-[5]. In this paper, we would only perform binary classification, classifying speakers into US accent or non-US accent. In general, a signal can be analyzed in time domain or in frequency domain. Usually, the analysis in time domain is an ill-posed problem due to the high dimensionality [6] [7]. Yet in frequency domain dimensionality reduction can be performed together with feature extraction via Mel-frequency cepstral coefficients [8] [9]. With the features being extracted from the raw signals, pattern recognition can be performed via multiple classifiers [10]-[12]. A block diagram of the process is shown in **Figure 1**.

The second and third sections explain the computation of MFCCs and different classifiers, focusing on discriminant analysis; support vector machine, and $k$-nearest neighbors, in detail. We describe the data being used in the paper in the fourth section and perform the result and discussion in the fifth section. Finally, we draw a brief conclusion and provide some future tasks in the last section.

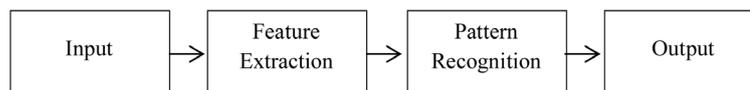

**Figure 1.** Block diagram of the process.

## 2. Feature Extraction via MFCCs

A voice signal in the time domain, which is simply a time series of the amplitude of the voice, is readily resulting in large number of variables. Consider a 5-second signal with a sampling rate of 8 kHz. It contains 40,000 entries which transforms into the same amount of variables in order to construct the data matrix being used in pattern recognition. Analysis with large number of variables generally will lead to intense computation and over-fitting. Fortunately, high dimensionality can be reduced through algorithms of feature extraction. In terms of voice signal, such an algorithm should be different from the common algorithms like principal component analysis, since we would like the algorithm not only reduce the dimensionality, but also retain the feature of the unique voice as much as possible. MFCC is a useful algorithm of performing feature extraction for voice signal [8].

The main idea of MFCC is to transform the signal from time domain to frequency domain and to map the transformed signal in hertz onto Mel-scale due to the fact that 1 kHz is a threshold of humans' hearing ability. Human ears are less sensitive to sound with frequency above this threshold. The calculation of MFCCs includes the following steps:
- Pre-emphasis filtering;
- Take the absolute value of the short time Fourier transformation using windowing;
- Warp to auditory frequency scale (Mel-scale);
- Take the discrete cosine transformation of the log-auditory-spectrum;
- Return the first q MFCCs.

Usually in a voice segment the spectrum has more energy at lower frequencies than at higher frequencies, but the signal to noise ratio (SNR) is lower at low frequencies. Pre-emphasis filtering, a special kind of finite impulse response (FIR), can be used to compensate this problem and provide more information by boosting the energy at higher frequencies. Let $x[n]$ be the raw signal at sample $n$, and $s[n]$ the signal after the high-pass filtering.

$$s[n] = x[n] - \alpha x[n-1], \ n = 1, 2, \cdots, N \tag{1}$$

where $\alpha$ is a parameter controlling how much is filtered and is often chosen between 0.95 and 1 in practice. **Figure 2** shows the difference between $x[n]$ and $s[n]$ in time domain.

The next step is to transform the signal from time domain to frequency domain by applying short time Fourier transformation together with a window function. One assumption of Fourier transformation is that the time series is stationary, which usually does not meet the situation when the signal is relatively long. Short time Fourier transformation assumes that the signal over a very short time period is at least nearly stationary thus able to be transformed to frequency domain. This can be done by

$$X_a[k] = \sum_{n=0}^{N-1} s[n] \cdot w_a[n] \cdot e^{-i2\pi kn/N} = \sum_{n=0}^{N-1} s[n] \cdot w_a[n] \cdot e^{-i\omega k}, \ 0 \le k < N, \tag{2}$$

where $w_a[n]$ is the window function, which is a zero valued function everywhere except inside the window $m$, and $i$ is the imaginary unit. Usually to keep the frames continuous, a Hamming window

$$w_a[n] = \alpha - \beta \cos\left(\frac{2\pi n}{N-1}\right), 0 \le n < N, \alpha = 0.54, \beta = 1 - \alpha = 0.46 \tag{3}$$

is preferred and the length of each frame is kept between 20 to 40 m. **Figure 3** demonstrates the effect of windowing with a frame length of 40 m.

A fact of human hearing ability is that we are more sensitive to sound between 20 and 1000 Hz. Thus it is less efficient to assign a signal the same scale at high frequencies as at lower frequencies. An adjustment can be

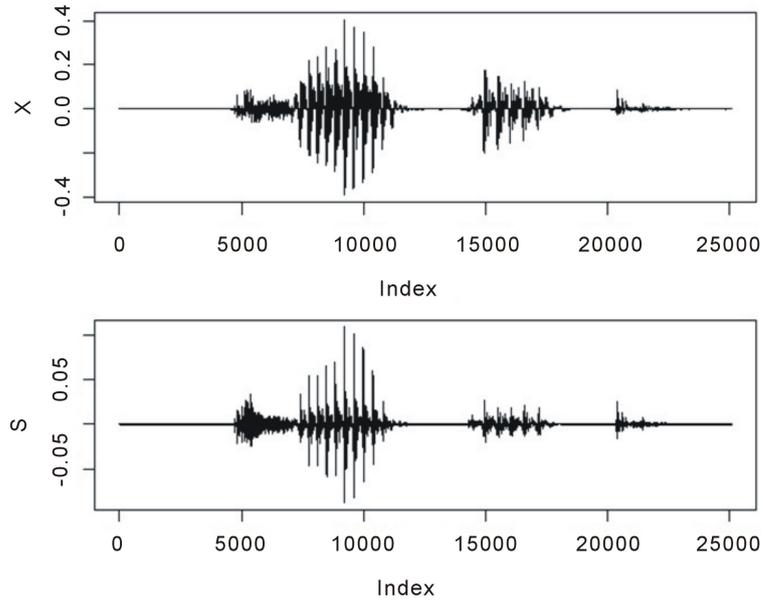

**Figure 2.** $x[n]$ and $s[n]$ in time domain.

made by mapping the data from Hertz-scale onto Mel-scale:

$$mel = \begin{cases} f, f \leq 1000 \\ 2595 \log_{10}\left(1 + \dfrac{f}{700}\right), f > 1000. \end{cases} \tag{4}$$

and its inverse is given by

$$f = \begin{cases} mel, mel \leq 1000 \\ 700\left(e^{mel/2595} - 1\right), mel > 1000 \end{cases} \tag{5}$$

A corresponding plot of this warping function is given in **Figure 4**.

Notice that when it covers a wide range on Hertz scale at high frequencies, it transforms onto Mel scale a much narrower range.

Given the STFT of a input window frame $x_a[k]$, we define a filterbank with $M$ filters ($m = 1, 2, \cdots, M$) that are linear on Mel scale but nonlinear on Hertz scale, where $m$ is triangular filter given by

$$M_m[k] = 1 - \left| \frac{k - \dfrac{N-1}{2}}{\dfrac{N-1}{2}} \right|, \tag{6}$$

where $N$ is the length of the filter. Notice again that these filters are linear on Mel scale and they need to be transformed back to Hertz scale. Thus we can then compute the log-energy of each filter as

$$S[m] = \ln\left[\sum_{k=0}^{N-1} \left|X_a[k]\right|^2 M_m[k]\right], 0 < m \leq M. \tag{7}$$

The Mel-frequency cepstrum coefficients are then the discrete cosine transform of the $M$ filter outputs:

$$c[q] = \sum_{m=0}^{M-1}\left[S[m]\cos\left(\dfrac{\pi q\left(m - \dfrac{1}{2}\right)}{M}\right)\right], 0 < m \leq M. \tag{8}$$

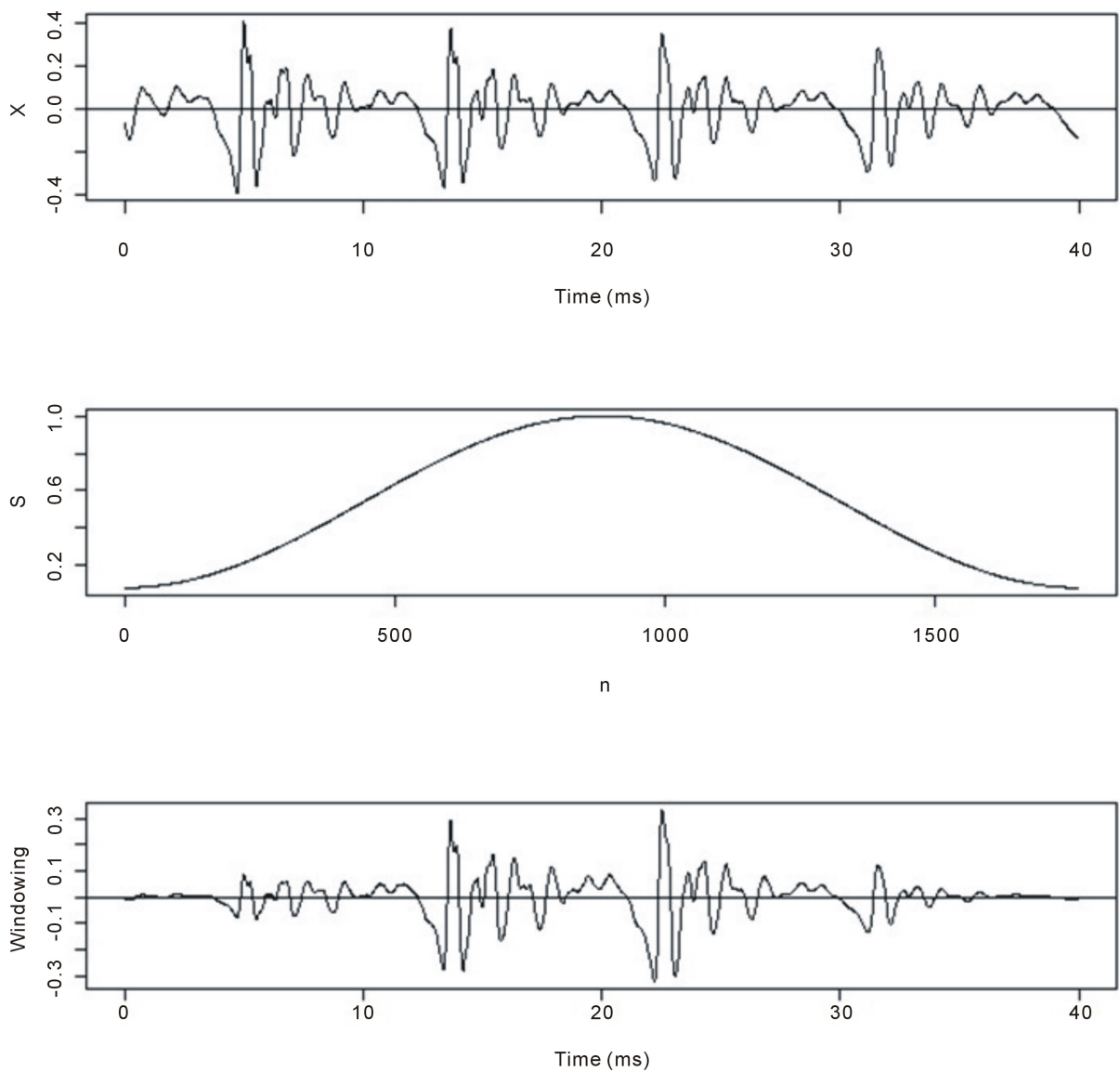

**Figure 3.** Effect of a Hamming window.

In practice, *M* is usually chosen between 24 and 40 and the first 13 MFCCs are computed. Also notice that for each signal, the MFCCs actually form a $n \times q$ matrix where n is the number of window frames and q is the number of MFCCs. If we are to pass the MFCC matrices to a vector based pattern recognition technique, these matrices have to be transformed or summarized to vectors. The simplest way of doing this is to take the mean values of each of the n column vectors.

## 3. Techniques in Classification

### 3.1. Discriminant Analysis

Discriminant analysis is one of the standard approaches to classification problems ([13], and [14]). Let the data matrix **X**, given every class *k* follow a Gaussian distribution

$$f_k(\mathbf{x}) = \frac{1}{\sqrt{(2\pi)^p |\Sigma_k|}} \exp\left(-\frac{1}{2}(\mathbf{x} - \mu_k)^T \Sigma_k^{-1} (\mathbf{x} - \mu_k)\right), \tag{9}$$

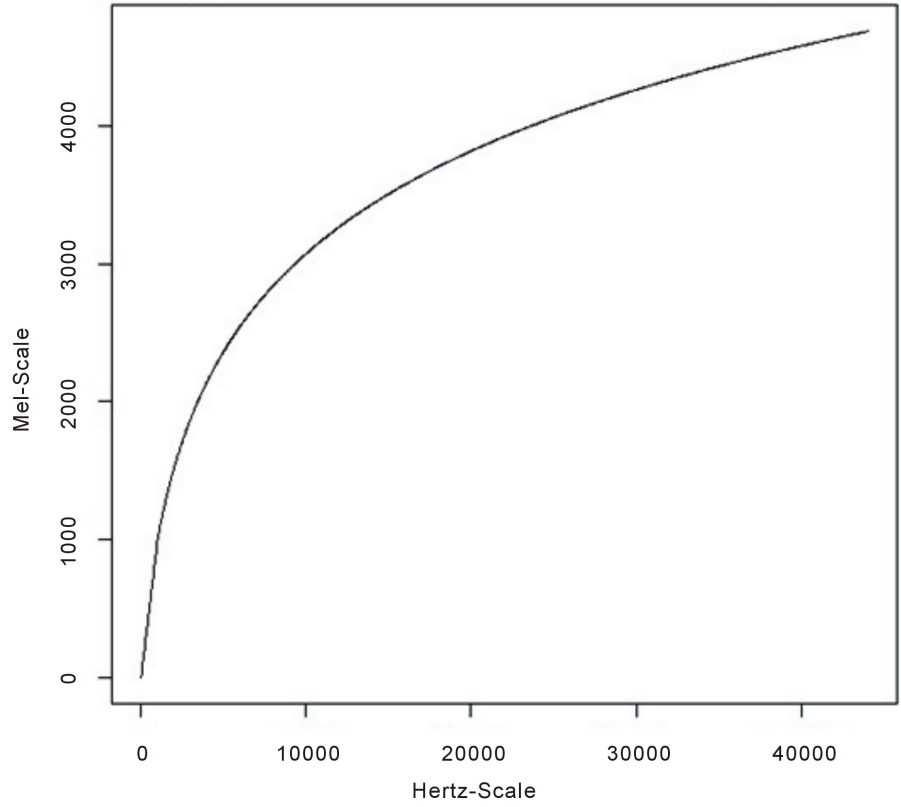

**Figure 4.** Relationship between Hertz-scale and Mel-scale.

where $p$ is the dimension and $\Sigma_k$ is the covariance matrix for class $k$. Both vector $x$ and mean vector $\mu_k$ are column vectors.

In linear discriminant analysis (LDA), we assume that the covariance matrices in all classes are the same, that is, $\Sigma_k = \Sigma, \forall k$. Based on Bayesian theory, we have

$$
\begin{aligned}
\hat{Y}(\mathbf{x}) &= \arg\max_k \Pr(Y = k | \mathbf{X} = \mathbf{x}) \\
&= \arg\max_k f_k(\mathbf{x}) \pi_k \\
&= \arg\max_k \left[ \mathbf{x}^T \Sigma^{-1} \mu_k - \frac{1}{2} \mu_k \Sigma^{-1} \mu_k + \log(\pi_k) \right],
\end{aligned}
\tag{10}
$$

where $\pi_k = \Pr(Y = k)$ is the prior probability. Define the linear discriminant function as

$$
\delta_k(\mathbf{x}) = \mathbf{x}^T \Sigma^{-1} \mu_k - \frac{1}{2} \mu_k \Sigma^{-1} \mu_k + \log(\pi_k)
\tag{11}
$$

Then

$$
\hat{Y}(\mathbf{x}) = \arg\max_k \delta_k(\mathbf{x})
\tag{12}
$$

In practice, $\mu_k$ and $\Sigma$ can be estimated by the sample mean and sample covariance.

Quadratic discriminant analysis (QDA) is almost the same as LDA, except that we no longer assume that the covariance matrix is the same for all classes. Thus, we have to estimate $\Sigma_k$ separately for each class $k$. The quadratic discriminant function is given by

$$
\delta_k(\mathbf{x}) = -\frac{1}{2} \log |\Sigma_k| - \frac{1}{2} (\mathbf{x} - \mu)^T \Sigma_k^{-1} (\mathbf{x} - \mu) + \log(\pi_k).
\tag{13}
$$

In both LDA and QDA, the classification rule is to search for the class $k$ which maximizes the discriminant function $\delta_k(\mathbf{x})$.

## 3.2. Support Vector Machines

The main idea of SVMs is to define a boundary between two classes by maximal separation of the closest observations. In practice, SVMs are powerful algorithm on binary classification tasks [13]-[15].

Given a data set $D = \left\{ (\mathbf{x}_i, y_i) \middle| \mathbf{x}_i \in R^p, y_i \in \{-1,1\} \right\}_{i=1}^n$, the general decision function of SVM is given by

$$\hat{f}_{SVM}(\mathbf{x}) = sign\left( \sum_i \hat{\alpha}_i y_i \Phi(\mathbf{x}_i) \cdot \Phi(\mathbf{x}) + \hat{b} \right) = sign\left( \sum_i \hat{\alpha}_i y_i K(\mathbf{x}_i, \mathbf{x}) + \hat{b} \right), \tag{14}$$

where $K(\mathbf{x}_i, \mathbf{x}) = \Phi(\mathbf{x}_i) \cdot \Phi(\mathbf{x})$ is called a kernel function and it is used for implicit nonlinear map. Common kernel functions are the Gaussian Radial Basis Function (RBF) kernel

$$K(\mathbf{x}_i, \mathbf{x}_j) = \exp\left( -\gamma \left\| \mathbf{x}_i - \mathbf{x}_j \right\|^2 \right) \tag{15}$$

and the polynomial kernel

$$K(\mathbf{x}_i, \mathbf{x}_j) = (\mathbf{x}_i \cdot \mathbf{x}_j + c)^d. \tag{16}$$

## 3.3. $k$-Nearest Neighbors

Comparing to the above two classifiers, the algorithm of $k$-nearest neighbors, which is a nonparametric method for classification, is more intuitive [13]-[15]. Given the training set $Tr = \left\{ (\mathbf{x}_i, y_i) \middle| \mathbf{x}_i \in R^p, y_i \in \{1,2,\cdots,S\} \right\}_{i=1}^n$ and a new data point $\mathbf{x}^*$, the distances between $\mathbf{x}^*$ and $\mathbf{x}_i$ are calculated based on some bivariate function $D(\cdot, \cdot)$. Then the distances are ranked in an increasing order and specify $V_k(\mathbf{x}^*) = \left\{ \mathbf{x}_i \middle| D(\mathbf{x}^*, \mathbf{x}_i) \leq D_{(k)} \right\}$, where $D_{(k)}$ the distance between the new point and the $k$th nearest neighbor. And the decision function can be defined as

$$\hat{f}_{kNN}(\mathbf{x}^*) = \underset{j \in \{1,2,\dots,S\}}{\arg\max} \left\{ \frac{1}{k} \sum_{i=1}^n I(y_i = j) \cdot I\left( \mathbf{x}_i \in V_k(\mathbf{x}^*) \right) \right\} \tag{17}$$

where $I(\cdot)$ is an indicator function.

By definition, two requirements are needed in order to perform the classification, a measurement of distance and a number $k$. Measurements like the Euclidean distance

$$D(\mathbf{x}_i, \mathbf{x}_j) = \sqrt{\sum_{l=1}^n (x_{il} - x_{jl})^2} \tag{18}$$

or the Manhattan distance

$$D(\mathbf{x}_i, \mathbf{x}_j) = \sum_{l=1}^n \left| x_{il} - x_{jl} \right| \tag{19}$$

are commonly used in the calculation of the distances. If a binary classification is performed, $k$ is better chosen to be an odd number.

## 4. Description of Data

A total of 330 signal data were collected from the voice of 22 speakers, 11 female and 11 male, from an internet source. Because of the method we used in collecting the data, there is no background noise in any sound tracks. 15 words were assigned to each voice and a demographic summary is given by the contingency **Table 1**.

Notice that the design is balanced in terms of accent but not gender. In this case, we would focus only on accent recognition.

Though the sound tracks have lengths of only around 1 second, with a sampling rate of 44,100 Hz, each sound track vector on the time domain has more than 30,000 entries. The response is given by

$$y_i = \begin{cases} 0, US \\ 1, non-US \end{cases} \qquad (20)$$

showing that a binary classification task is performed.

## 5. Comparison of Different Classifiers

The MFCCs were computed for each sound track and the mean vectors were passed to different classifiers. For discriminant analysis, both LDA and QDA were implemented. For SVMs, both RBF and 2nd degree polynomial kernels were used and then were compared to each other. For $k$-NN, $k$ was chosen to be 3 in a preliminary analysis. In order to approximate the true prediction ability of the model, a holdout cross-validation of size 500 was done based on stratified random sampling. The precision for each prediction is simply the ratio between the correct prediction, which is the summation of true positive (*TP*) and true negative (*TN*), and the total number of sound tracks (*N*). And the overall prediction accuracy is the average accuracy of the cross-validation of size $m = 500$.

$$\text{Accuracy} = \frac{\sum \left\{ 1 - \frac{TP + TN}{N} \right\}}{m} \qquad (21)$$

**Table 2** gives the average accuracy of the different classifiers at each one of the 5 levels of MFCCs.

**Figure 5** is a corresponding plot of this table, showing a comparison of prediction accuracy of the 5 classifiers. The graph provides the average accuracy of a cross-validation of size 500. It is of interest to see that apart from LDA, which is less competitive comparing to the others, all the other four techniques have similar results. The prediction accuracy increases together with the number of MFCCs, but as the number of MFCCs is beyond 30, the prediction accuracy does not increase as obvious as that when the number of MFCCs increases from 12 to 26. Also, $k$-NN, regardless of its intuitive algorithm, yields more accurate results than the other classifiers.

Moreover, the time being used in the computation is given in **Table 3**.

**Figure 6** is a corresponding plot if **Table 3**. Notice that $k$-NN uses least time, around 1 second, to finish the computation of cross-validation while the other techniques use much more time. This is not surprising, since $k$-NN algorithm, unlike the other classifiers, does not intend to build models or estimate parameters.

## 6. Conclusions

We have demonstrated in this paper that pattern recognition of signals can be performed through different classifiers combined with the MFCC features. In terms of feature extraction, the number of MFCCs below 30 seems to be a reasonable choice. With this number increasing, we may face problems caused by high dimensionality although the information becomes richer when more MFCCs are involved in the computation. Comparing different classifiers, $k$-nearest neighbor is the most powerful tool in this task. Not only does it yield the most accurate prediction results, it also uses the least time in the performance.

**Table 1.** A demographic summary of speakers.

| Accent | Gender | | |
| --- | --- | --- | --- |
| | Female | Male | Total |
| US | 90 | 75 | 165 |
| Non-US | 90 | 75 | 165 |
| Total | 180 | 150 | 330 |

**Table 2.** Average accuracy of classifiers.

| # MFCCs | LDA | QDA | SVM (RBF) | SVM (PLY) | $k$-NN |
| --- | --- | --- | --- | --- | --- |
| 12 | 0.7353 | 0.8112 | 0.8208 | 0.8097 | 0.8548 |
| 19 | 0.7503 | 0.8647 | 0.8507 | 0.8851 | 0.9098 |
| 26 | 0.8063 | 0.9224 | 0.9080 | 0.9379 | 0.9398 |
| 33 | 0.8319 | 0.9543 | 0.9352 | 0.9509 | 0.9586 |
| 39 | 0.8260 | 0.9383 | 0.9223 | 0.9438 | 0.9605 |

**Table 3.** Comparison of computation time (in seconds) of classifiers.

| # MFCCs | LDA | QDA | SVM (RBF) | SVM (PLY) | k-NN |
|---|---|---|---|---|---|
| 12 | 7.36 | 7.26 | 11.75 | 10.58 | 0.64 |
| 19 | 10.03 | 9.63 | 13.37 | 9.55 | 0.85 |
| 26 | 12.12 | 11.94 | 15.08 | 11.51 | 1.10 |
| 33 | 14.87 | 14.46 | 16.02 | 12.36 | 1.03 |
| 39 | 18.21 | 16.78 | 16.54 | 12.64 | 1.15 |

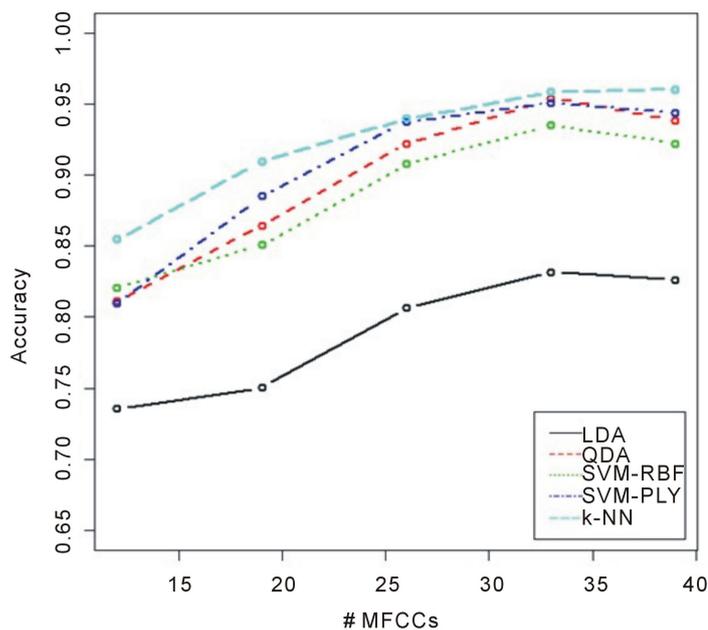

**Figure 5.** Comparison of accuracy.

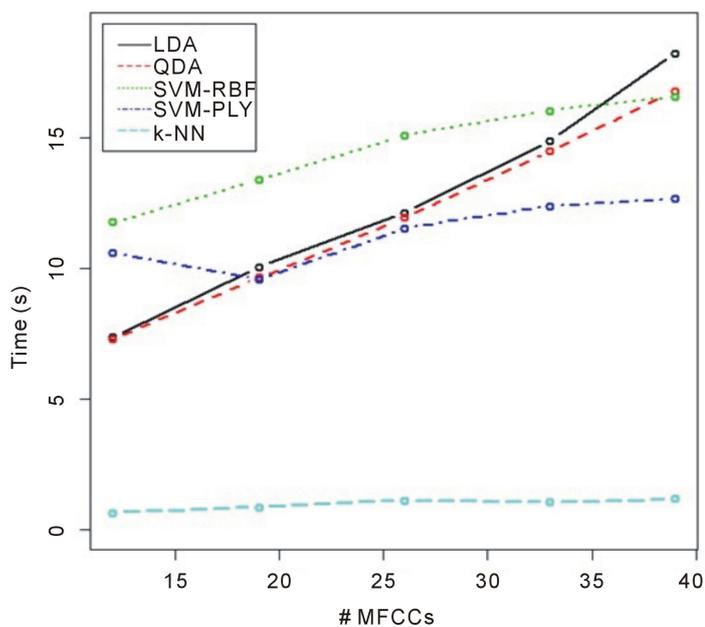

**Figure 6.** Comparison of time.

In this paper, we only considered the mean vectors of MFCCs matrices for simplicity, but alternative methods can be taken into account to generate the input for pattern recognition. For instance, the standard deviations of each MFCC can be used together with the mean values, or each coefficient can be modelled as a Gaussian mixture. Also, feature extraction via MFCCs is not as powerful when the signal contains significant noise since MFCCs provide detailed information of the raw signal. In such cases, alternative algorithms, usually much more complex, should be preferred rather than MFCCs.

## Acknowledgements

Ernest Fokoué wishes to express his heartfelt gratitude and infinite thanks to Our Lady of Perpetual Help for Her ever-present support and guidance, especially for the uninterrupted flow of inspiration received through Her most powerful intercession.

Scientific Research Publishing (SCIRP) is one of the largest Open Access journal publishers. It is currently publishing more than 200 open access, online, peer-reviewed journals covering a wide range of academic disciplines. SCIRP serves the worldwide academic communities and contributes to the progress and application of science with its publication.

Other selected journals from SCIRP are listed as below. Submit your manuscript to us via either submit@scirp.org or Online Submission Portal.

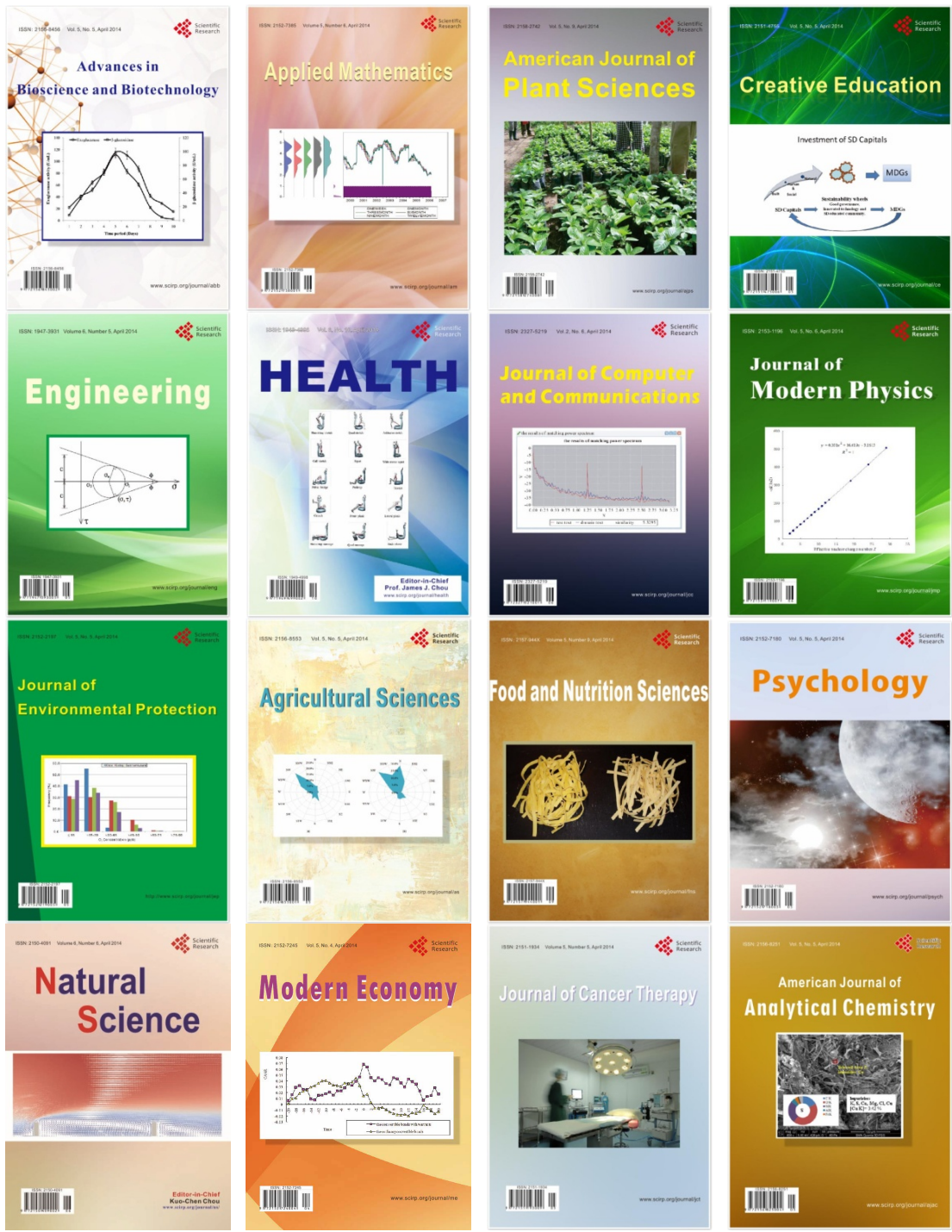